\documentclass[preprint,nopreprintnumbers,titlepage,amsmath,amssymb,aps,prl]{revtex4-2}

\usepackage{graphicx}
\usepackage{dcolumn}
\usepackage{bm}
\usepackage{hyperref}

\usepackage{braket}
\usepackage{lipsum}
\usepackage{alltt}
\usepackage{multirow}
\usepackage{stackrel}

\makeatletter
\allowdisplaybreaks[1]

\newcommand{\e}{{\rm e}}
\newcommand{\tr}{{\mathrm{tr}}}

\newcommand{\dd}{\,\mathrm{d}}

\begin{document}
\title{Proof of Sarkar-Kumar's Conjectures on Average Entanglement Entropies over the Bures-Hall Ensemble}

\author{Lu Wei}
\email{luwe@umich.edu}
\affiliation{Department of Electrical and Computer Engineering \\ University of Michigan - Dearborn, MI 48128, USA}
\date{\today}

\begin{abstract}
Sarkar and Kumar recently conjectured~[J. Phys. A: Math. Theor. $\bm{52}$, 295203 (2019)] that for a bipartite system of Hilbert dimension $mn$, the mean values of quantum purity and von Neumann entropy of a subsystem of dimension $m\leq n$ over the Bures-Hall measure are given by
\begin{equation*}
\frac{2n(2n+m)-m^{2}+1}{2n(2mn-m^2+2)}
\end{equation*}
and
\begin{equation*}
\psi_{0}\left(mn-\frac{m^2}{2}+1\right)-\psi_{0}\left(n+\frac{1}{2}\right),
\end{equation*}
respectively, where $\psi_{0}(\cdot)$ is the digamma function. We prove the above conjectured formulas in this work. A key ingredient of the proofs is Forrester and Kieburg's discovery on the connection between the Bures-Hall ensemble and the Cauchy-Laguerre biorthogonal ensemble studied by Bertola, Gekhtman, and Szmigielski.
\end{abstract}

\maketitle


\section{Introduction and the Conjectures}
Consider a composite quantum system that consists of two subsystems $A$ and $B$ of Hilbert space dimensions $m$ and $n$, respectively. The Hilbert space $\mathcal{H}_{A+B}$ of the composite system is given by the tensor product of the Hilbert spaces of the subsystems, $\mathcal{H}_{A+B}=\mathcal{H}_{A}\otimes\mathcal{H}_{B}$. Define a state of the composite system as a linear combination of the random coefficients $z_{i,j}$ and the complete basis $\left\{\Ket{i^{A}}\right\}$ and $\left\{\Ket{j^{B}}\right\}$ of $\mathcal{H}_{A}$ and $\mathcal{H}_{B}$,
\begin{equation}\label{eq:S0}
\Ket{\psi}=\sum_{i=1}^{m}\sum_{j=1}^{n}z_{i,j}\Ket{i^{A}}\otimes\Ket{j^{B}},
\end{equation}
we then consider a superposition of the state Eq.~(\ref{eq:S0}) as
\begin{equation}\label{eq:SB}
\Ket{\varphi}=\Ket{\psi}+\left(\mathbf{U}\otimes I_{m}\right)\Ket{\psi},
\end{equation}
where $\mathbf{U}$ is a unitary matrix taken from a certain measure~\cite{Sarkar19}. The corresponding density matrix is
\begin{equation}\label{eq:rho}
\rho=\Ket{\varphi}\Bra{\varphi},
\end{equation}
which has the natural probability constraint
\begin{equation}\label{eq:del}
\tr(\rho)=1.
\end{equation}
We assume without loss of generality that $m\leq n$. The reduced density matrix $\rho_{A}$ of the smaller subsystem $A$ is computed by partial tracing of the full density matrix Eq.~(\ref{eq:rho}) over the other subsystem $B$ (interpreted as the environment) as
\begin{equation}\label{eq:rhoB}
\rho_{A}=\tr_{B}\rho.
\end{equation}
The resulting density of eigenvalues of $\rho_{A}$ ($\lambda_{i}\in[0,1]$, $i=1,\dots,m$) is the (generalized) Bures-Hall measure~\cite{Sarkar19}
\begin{equation}\label{eq:BH}
f\left(\bm{\lambda}\right)=\frac{1}{c}~\delta\left(1-\sum_{i=1}^{m}\lambda_{i}\right)\prod_{1\leq i<j\leq m}\frac{\left(\lambda_{i}-\lambda_{j}\right)^{2}}{\lambda_{i}+\lambda_{j}}\prod_{i=1}^{m}\lambda_{i}^{\alpha},
\end{equation}
where the parameter $\alpha$ takes half-integer values
\begin{equation}\label{eq:aBH}
\alpha=n-m-\frac{1}{2},
\end{equation}
and the constant $c$ is
\begin{equation}\label{eq:cBH}
c=\frac{2^{-m(m+2\alpha)}\pi^{m/2}}{\Gamma\left(m(m+2\alpha+1)/2\right)}\prod_{i=1}^{m}\frac{\Gamma(i+1)\Gamma(i+2\alpha+1)}{\Gamma(i+\alpha+1/2)}.
\end{equation}
In Eq.~(\ref{eq:BH}), the presence of the Dirac delta function $\delta(\cdot)$ reflects the constraint Eq.~(\ref{eq:del}). The Bures-Hall measure enjoys the property that, without any prior knowledge on a density matrix, the optimal way to estimate the density matrix is to generate a state at random with respect to the Bures-Hall measure~\cite{Osipov10}. Thus, it is often used as a prior distribution (Bures prior) in reconstructing quantum states from measurements.

The degree of entanglement of subsystems can be measured by the entanglement entropies, which are functions of eigenvalues of the reduced density matrix Eq.~(\ref{eq:rhoB}). An entanglement entropy should monotonically change from the separable state ($\lambda_{1}=1$, $\lambda_{2}=\dots=\lambda_{m}=0$) to the maximally-entangled state ($\lambda_{1}=\lambda_{2}=\dots\lambda_{m}=1/m$). A standard one is the quantum purity~\cite{BZ06}
\begin{equation}\label{eq:P}
S_{\text{P}}=\tr\left(\rho_{A}^{2}\right)=\sum_{i=1}^{m}\lambda_{i}^{2},
\end{equation}
supported in $S_{\text{P}}\in[1/m,1]$, which measures how far a state is from a pure state $\rho_{A}^{2}=\rho_{A}$ that corresponds to $S_{\text{P}}=1$. Quantum purity Eq.~(\ref{eq:P}) is an example of polynomial entropies, whereas a well-known non-polynomial entropy is the von Neumann entropy~\cite{BZ06}
\begin{equation}\label{eq:vN}
S_{\text{vN}}=-\tr\left(\rho_{A}\ln\rho_{A}\right)=-\sum_{i=1}^{m}\lambda_{i}\ln\lambda_{i},
\end{equation}
supported in $S_{\text{vN}}\in[0,\ln{m}]$, which achieves the separable state and maximally-entangled state when $S_{\text{vN}}=0$ and when $S_{\text{vN}}=\ln{m}$, respectively.

Statistical information of entropies is encoded through their moments: the first moment (average value) implies the typical behavior of entanglement and the higher moments specify fluctuation around the typical values. For the Hilbert-Schmidt measure~\cite{BZ06}, that corresponds to the density without the interaction term
\begin{equation}
\prod_{1\leq i<j\leq m}(\lambda_{i}+\lambda_{j})
\end{equation}
in Eq.~(\ref{eq:BH}), the moments of quantum purity~\cite{Lubkin78,Giraud07} and von Neumann entropy~\cite{Page93,Foong94,Ruiz95,VPO16,Wei17,Wei20} have been well-investigated~\footnote{For a comprehensive treatment of the density matrix formulism including the discussed measures and entropies, we refer readers to Ref.~\cite{BZ06} and references therein.}. However, knowledge on the behavior of entanglement entropies over the Bures-Hall measure is quite limited. In the special case of equal subsystem dimensions $m=n$, i.e., $\alpha=1/2$ in Eq.~(\ref{eq:aBH}), the resulting moments of purity were derived in Refs.~\cite{Sommers04,Osipov10}. For arbitrary subsystem dimensions $m\leq n$, Sarkar and Kumar recently conjectured~\cite[Eqs.~(61) and~(59)]{Sarkar19} that the average quantum purity and the average von Neumann entropy are given by (notice the notation difference here and in Ref.~\cite{Sarkar19})
\begin{equation}\label{eq:Pm}
\mathbb{E}_{f}\!\left[S_{\text{P}}\right]=\frac{2n(2n+m)-m^{2}+1}{2n(2mn-m^2+2)}
\end{equation}
and
\begin{equation}\label{eq:vNm}
\mathbb{E}_{f}\!\left[S_{\text{vN}}\right]=\psi_{0}\left(mn-\frac{m^2}{2}+1\right)-\psi_{0}\left(n+\frac{1}{2}\right),
\end{equation}
respectively, where the expectations $\mathbb{E}_{f}\!\left[\cdot\right]$ are taken over the Bures-Hall ensemble Eq.~(\ref{eq:BH}). Here,
$\psi_{0}(x)=\dd\ln\Gamma(x)/\dd x$ is the digamma function (Psi function)~\cite{Prudnikov86} and for a positive integer $l$,
\begin{subequations}
\begin{eqnarray}
\psi_{0}(l)&=&-\gamma+\sum_{k=1}^{l-1}\frac{1}{k},\\
\psi_{0}\left(l+\frac{1}{2}\right)&=&-\gamma-2\ln2+2\sum_{k=0}^{l-1}\frac{1}{2k+1},
\end{eqnarray}
\end{subequations}
where $\gamma\approx0.5772$ is the Euler's constant. In the rest of the paper, we show that the conjectured formulas~(\ref{eq:Pm}) and~(\ref{eq:vNm}) are indeed correct.

\section{Average Entropies over Bures-Hall Ensemble}
\subsection{Moment Relations}
The first step is a rather standard calculation, briefly outlined below (see also, e.g., Refs.~\cite{Sarkar19,Osipov10,Page93,Ruiz95,Wei17,Wei20}), that relates the moment computation over an ensemble with the constraint $\delta\left(1-\sum_{i=1}^{m}\lambda_{i}\right)$ to a one without. As will be seen, the corresponding unconstrained ensemble of the Bures-Hall ensemble Eq.~(\ref{eq:BH}) is~\cite{Sarkar19}
\begin{equation}\label{eq:BHu}
h\left(\bm{x}\right)=\frac{1}{c'}\prod_{1\leq i<j\leq m}\frac{\left(x_{i}-x_{j}\right)^{2}}{x_{i}+x_{j}}\prod_{i=1}^{m}x_{i}^{\alpha}\e^{-x_{i}},
\end{equation}
where $x_{i}\in[0,\infty)$, $i=1,\dots,m$, and the constant $c'$ is related to the constant Eq.~(\ref{eq:cBH}) by
\begin{equation}\label{eq:cBHu}
c'=c~\Gamma\left(m(m+2\alpha+1)/2\right).
\end{equation}
Despite being only interested in the half-integer values of $\alpha$ in Eq.~(\ref{eq:aBH}), the following results, in particular Eqs.~(\ref{eq:aPm}) and~(\ref{eq:avNm}), are valid for $\alpha>-1$ that the density Eq.~(\ref{eq:BHu}) is defined. We start with finding the first moment relation for the von Neumann entropy, where, by multiplying an auxiliary integral over a gamma density, one has
\begin{equation}\label{eq:mc}
\mathbb{E}_{f}\!\left[S_{\text{vN}}\right] = \int_{0}^{\infty}\frac{\e^{-\theta}\theta^{d-1}}{\Gamma(d)}\dd\theta\int_{\bm{\lambda}}S_{\text{vN}}f\left(\bm{\lambda}\right)\prod_{i=1}^{m}\dd\lambda_{i}.
\end{equation}
Inserting the the change of variables
\begin{equation}\label{eq:cv}
\lambda_{i}=\frac{x_{i}}{\theta},~~~~i=1,\ldots,m,
\end{equation}
into Eq.~(\ref{eq:mc}), some simplification leads to
\begin{eqnarray}\label{eq:mc1}
\mathbb{E}_{f}\!\left[S_{\text{vN}}\right]&=&\psi_{0}(d)-\frac{c^{-1}}{\Gamma(d)}\int_{\bm{x}}\prod_{1\leq i<j\leq m}\frac{\left(x_{i}-x_{j}\right)^{2}}{x_{i}+x_{j}}\prod_{i=1}^{m}x_{i}^{\alpha} \nonumber \\
&&\times\!\int_{0}^{\infty}\!\!\!\!\e^{-\theta}\theta^{d-m(m+1)/2-\alpha m-1}\delta\!\left(\theta-\sum_{i=1}^{m}x_{i}\right)\!\!\dd\theta\prod_{i=1}^{m}\!\dd\lambda_{i},
\end{eqnarray}
where we also used
\begin{equation}\label{eq:1eln}
\int_{0}^{\infty}\!\!\e^{-\theta}\theta^{d-1}\ln{\theta}\dd \theta=\Gamma(d)\psi_{0}(d),~~~~\Re(d)>0.
\end{equation}
By setting $d=m(m+1)/2+\alpha m+1$, the integral over $\theta$ in Eq.~(\ref{eq:mc1}) can be conveniently evaluated that leads to the first moment relation as
\begin{equation}\label{eq:mc2}
\mathbb{E}_{f}\!\left[S_{\text{vN}}\right]=\psi_{0}\left(\frac{m(m+1)}{2}+\alpha m+1\right)-\frac{2}{m(m+2\alpha+1)}\mathbb{E}_{h}\!\left[T_{\text{vN}}\right],
\end{equation}
where we used the identity Eq.~(\ref{eq:cBHu}), and the random variable
\begin{equation}\label{eq:TvN}
T_{\text{vN}}=\sum_{i=1}^{m}x_{i}\ln x_{i}
\end{equation}
is understood as the induced von Neumann entropy over the unconstrained ensemble Eq.~(\ref{eq:BHu}). In a similar but more straightforward manner, the first moment relation for quantum purity is obtained as (see also Ref.~\cite{Sarkar19})
\begin{equation}\label{eq:mc3}
\mathbb{E}_{f}\!\left[S_{\text{P}}\right]=\frac{4}{m(m+2\alpha+1)\left(m^2+2\alpha m+m+2\right)}\mathbb{E}_{h}\!\left[T_{\text{P}}\right],
\end{equation}
where $T_{\text{P}}$ is the induced purity
\begin{equation}\label{eq:TP}
T_{\text{P}}=\sum_{i=1}^{m}x_{i}^{2}
\end{equation}
over the unconstrained ensemble Eq.~(\ref{eq:BHu}).

Proving Eq.~(\ref{eq:Pm}) and Eq.~(\ref{eq:vNm}) now boils down to computing the induced first moments $\mathbb{E}_{h}\!\left[T_{\text{P}}\right]$ in Eq.~(\ref{eq:mc3}) and $\mathbb{E}_{h}\!\left[T_{\text{vN}}\right]$ in Eq.~(\ref{eq:mc2}), respectively. Computing these average values requires the one-point correlation function~\cite{Mehta,Forrester}, i.e., the density of an arbitrary eigenvalue, of the unconstrained Bures-Hall ensemble Eq.~(\ref{eq:BHu}). In fact, its $k$-point correlation function was recently derived in Ref.~\cite{Forrester16}, which is written in terms of the correlation functions of the Cauchy-Laguerre biorthogonal ensemble~\cite{Bertola14}. In particular, the needed an arbitrary eigenvalue density of the unconstrained ensemble Eq.~(\ref{eq:BHu}) is~\cite{Forrester16}
\begin{equation}\label{eq:one}
h_{1}(x)=\frac{1}{2m}\big(G_{\alpha}(x)+G_{\alpha+1}(x)\big),
\end{equation}
where we denote
\begin{equation}\label{eq:GG}
G_{q}(x)=\int_{0}^{1}G_{2,3}^{1,1}(q|tx)G_{2,3}^{2,1}(q|tx)\dd t
\end{equation}
with
\begin{eqnarray}
G_{2,3}^{1,1}\left(q|x\right)&=&G_{2,3}^{1,1}\left(\begin{array}{c} -m; m+2\alpha+1 \\ 2\alpha+1; 0, q \end{array}\Big|x\Big.\right),\label{eq:G1}\\
G_{2,3}^{2,1}\left(q|x\right)&=&G_{2,3}^{2,1}\left(\begin{array}{c} -m-2\alpha-1; m \\ 0, -q; -2\alpha-1 \end{array}\Big|x\Big.\right)\label{eq:G2}
\end{eqnarray}
further denoting some Meijer G-functions~\cite{Prudnikov86}. In general, the Meijer G-function is defined by the following contour integral~\cite{Prudnikov86}
\begin{eqnarray}\label{eq:MG}
&&G_{p,q}^{m,n}\left(\begin{array}{c} a_{1},\ldots,a_{n}; a_{n+1},\ldots,a_{p} \\ b_{1},\ldots,b_{m}; b_{m+1},\ldots,b_{q} \end{array}\Big|x\Big.\right)\nonumber\\
&&=\frac{1}{2\pi\imath}\int_{\mathcal{L}}{\frac{\prod_{j=1}^m\Gamma\left(b_j+s\right)\prod_{j=1}^n\Gamma\left(1-a_j-s\right)x^{-s}}{\prod_{j=n+1}^p \Gamma\left(a_{j}+s\right)\prod_{j=m+1}^q\Gamma\left(1-b_j-s\right)}}\dd s,
\end{eqnarray}
where the contour $\mathcal{L}$ separates the poles of $\Gamma\left(1-a_j-s\right)$ from the poles of $\Gamma\left(b_j+s\right)$.

It will become clear that as intermediate steps to obtain $\mathbb{E}_{h}\!\left[T_{\text{P}}\right]$ and $\mathbb{E}_{h}\!\left[T_{\text{vN}}\right]$, we need to compute the integral below involving the Meijer G-functions Eqs.~(\ref{eq:G1}) and~(\ref{eq:G2})
\begin{equation}\label{eq:Ib}
I_{q}^{(\beta)}(t)=\int_{0}^{\infty}x^{\beta}G_{2,3}^{1,1}(q|tx)G_{2,3}^{2,1}(q|tx)\dd x,~~~~t>0,
\end{equation}
for $\beta=0,1,2$, as well as its derivative for $\beta=1$,
\begin{equation}\label{eq:IH}
H_{q}(t)=\frac{\dd}{\dd\beta}I_{q}^{(\beta)}(t)\Big|_{\beta=1},
\end{equation}
where $q$ will take the values $\alpha$ and $\alpha+1$ in both Eqs.~(\ref{eq:Ib}) and~(\ref{eq:IH}). To compute Eq.~(\ref{eq:Ib}), we use the fact that the Meijer G-function Eq.~(\ref{eq:G1}) can be written as a terminating hypergeometric function~\cite{Prudnikov86} (see also Refs.~\cite{Bertola14,Forrester16})
\begin{eqnarray}
G_{2,3}^{1,1}\left(q|tx\right)&=&\frac{\Gamma(m+2\alpha+2)}{\Gamma(m)\Gamma(2\alpha+2)\Gamma(2\alpha+2-q)}(tx)^{2\alpha+1}~\!_{2}F_{2}\left(\begin{array}{c} 1-m, m+2\alpha+2 \\ 2\alpha+2, 2\alpha+2-q \end{array}\Big|tx\Big.\right) \\
&=&\frac{\Gamma(m+2\alpha+2)}{\Gamma(m)\Gamma(2\alpha+2)\Gamma(2\alpha+2-q)}(tx)^{2\alpha+1}\sum_{k=0}^{m-1}\frac{(1-m)_{k}(m+2\alpha+2)_{k}(tx)^{k}}{(2\alpha+2)_{k}(2\alpha+2-q)_{k}k!},\label{eq:M2H}
\end{eqnarray}
where $(a)_{n}=\Gamma(a+n)/\Gamma(a)$ is the Pochhammer symbol. Inserting Eq.~(\ref{eq:M2H}) into Eq.~(\ref{eq:Ib}), the integral can now be evaluated by using the Mellin transform of the Meijer G-function~\cite{Prudnikov86} (cf. Eq.~(\ref{eq:MG}))
\begin{eqnarray}\label{eq:iMG}
&&\int_{0}^{\infty}x^{s-1}G_{p,q}^{m,n}\left(\begin{array}{c} a_{1},\ldots,a_{n}; a_{n+1},\ldots,a_{p} \\ b_{1},\ldots,b_{m}; b_{m+1},\ldots,b_{q} \end{array}\Big|\eta x\Big.\right)\dd x \nonumber \\
&&=\frac{\eta^{-s}\prod_{j=1}^m\Gamma\left(b_j+s\right)\prod_{j=1}^{n}\Gamma\left(1-a_j-s\right)}{\prod_{j=n+1}^{p}\Gamma\left(a_{j}+s\right)\prod_{j=m+1}^q\Gamma\left(1-b_j-s\right)}
\end{eqnarray}
valid for $\Re(s)>-\min_{1\leq j\leq m}\Re(b_{j})$ and $\eta>0$, as
\begin{equation}\label{eq:Ibs0}
I_{q}^{(\beta)}(t)=t^{-\beta-1}I_{q}^{(\beta)},
\end{equation}
where $I_{q}^{(\beta)}$ denotes the $t$ independent part
\begin{eqnarray}\label{eq:Ibs}
I_{q}^{(\beta)}&=&\sum_{k=0}^{m-1}\frac{(-1)^{k+m}\Gamma(k+2\alpha+m+2)\Gamma(k+\beta+1)}{\Gamma(k+2\alpha+2)\Gamma(k+2\alpha+2-q)\Gamma(m-k)k!}\nonumber\\
&&\times\frac{\Gamma(k+\beta+2\alpha+2)\Gamma(k+\beta+2\alpha+2-q)}{\Gamma(k+\beta+2\alpha+m+2)\Gamma(k+\beta-m+1)}.
\end{eqnarray}
In obtaining Eq.~(\ref{eq:Ibs}), we also used the result of gamma function of negative arguments
\begin{equation}\label{eq:ig}
\Gamma(-l+\epsilon)=\frac{(-1)^{l}}{l!\epsilon}\left(1+o\left(\epsilon\right)\right)
\end{equation}
to resolve some indeterminacy by taking the limit $\epsilon\to0$. Since the $q$ dependent term in Eq.~(\ref{eq:Ibs}) is $(k+2\alpha+2-q)_{\beta}$, $I_{q}^{(\beta)}(t)$ becomes a $\beta$-th degree polynomial in $q$ for a non-negative integer $\beta$. The needed cases when $\beta=0,1,2$ can now be directly obtained as
\begin{subequations}
\begin{eqnarray}
I_{q}^{(0)}(t)&=&0, \label{eq:b0} \\
I_{q}^{(1)}(t)&=&-\frac{m(m+2\alpha+1)(m+2\alpha+1-q)}{2m+2\alpha+1}t^{-2}, \label{eq:b1} \\
I_{q}^{(2)}(t)&=&-\frac{m(m+2\alpha+1)(m+2\alpha+1-q)}{2(m+\alpha)(m+\alpha+1)(2m+2\alpha+1)}\Big((m+2\alpha+1)\left(5m^2+8\alpha m+4m+4\alpha^2+4\alpha\right) \nonumber \\
&&-(3m^2+6\alpha m+3m+4\alpha^2+4\alpha)q\Big)t^{-3}, \label{eq:b2}
\end{eqnarray}
\end{subequations}
where the non-zero contribution in Eq.~(\ref{eq:Ibs}) for $\beta=1$ and $\beta=2$ is from the terms $k=m-1$ and $k=m-2,m-1$, respectively. As a consequence of Eq.~(\ref{eq:b0}), the integral Eq.~(\ref{eq:GG}) can be also represented, by the symmetry of Eq.~(\ref{eq:Ib}) in $t$ and $x$ when $\beta=0$, as
\begin{equation}\label{eq:01}
G_{q}(x)=-\int_{1}^{\infty}G_{2,3}^{1,1}(q|tx)G_{2,3}^{2,1}(q|tx)\dd t.
\end{equation}
To evaluate Eq.~(\ref{eq:IH}), we first notice from Eqs.~(\ref{eq:Ibs0}) and~(\ref{eq:Ib}) that
\begin{equation}\label{eq:IH0}
H_{q}(t)=t^{-2}H_{q}-I_{q}^{(1)}(t)\ln t,
\end{equation}
where $I_{q}^{(1)}(t)$ has been computed in Eq.~(\ref{eq:b1}), and $H_{q}$ similarly denotes (cf. Eq.~(\ref{eq:IH}))
\begin{equation}\label{eq:IHsd}
H_{q}=\frac{\dd}{\dd\beta}I_{q}^{(\beta)}\Big|_{\beta=1}.
\end{equation}
By invoking Eq.~(\ref{eq:ig}) and the limiting behavior of digamma function
\begin{equation}\label{eq:idg}
\psi_{0}(-l+\epsilon)=-\frac{1}{\epsilon}\left(1+o\left(\epsilon\right)\right)
\end{equation}
to resolve an indeterminacy, $H_{q}$ is obtained as
\begin{eqnarray}\label{eq:IHs}
H_{q}&=&-\frac{m(m+2\alpha+1)(m+2\alpha+1-q)}{2m+2\alpha+1}\big(\psi_{0}(m+1)+\psi_{0}(m+2\alpha+2)+\psi_{0}(m+2\alpha+2-q) \nonumber \\
&&\!\!-\psi_{0}(2m+2\alpha+2)-\psi_{0}(1)\big)+\sum_{k=0}^{m-2}\frac{(k+1)(k+2\alpha+2)(k+2\alpha+2-q)}{(m-k-1)(k+m+2\alpha+2)}.
\end{eqnarray}
With the help of the identity
\begin{equation}\label{eq:dgs}
\psi_{0}(l+n)=\psi_{0}(l)+\sum_{k=0}^{n-1}\frac{1}{l+k},
\end{equation}
further simplification of Eq.~(\ref{eq:IHs}) gives
\begin{eqnarray}\label{eq:IHf}
H_{q}&=&\frac{m(m+2\alpha+1)}{2m+2\alpha+1}\bigg(\frac{a_{1}+2a_{2}q}{2(m+2\alpha+1)(2m+2\alpha+1)}+(2\alpha+1-2q)(\psi_{0}(2m+2\alpha+2)\nonumber\\
&&-\psi_{0}(m+2\alpha+2))-(m+2\alpha+1-q)\psi_{0}(m+2\alpha+1-q)\bigg),
\end{eqnarray}
where we denote
\begin{subequations}
\begin{eqnarray}
a_{1}&=&-4m^3-24\alpha m^2-14m^2-36\alpha^2m-40\alpha m-11m-16\alpha^3-28\alpha^2-16\alpha-3, \\
a_{2}&=&4m^2+8\alpha m+3m+4\alpha^2+4\alpha+1.
\end{eqnarray}
\end{subequations}
With the above preparations, we now derive expressions for $\mathbb{E}_{h}\!\left[T_{\text{P}}\right]$ in Eq.~(\ref{eq:mc3}) and $\mathbb{E}_{h}\!\left[T_{\text{vN}}\right]$ in Eq.~(\ref{eq:mc2}).

\subsection{Average Quantum Purity}
By definition, the mean value $\mathbb{E}_{h}\!\left[T_{\text{P}}\right]$ is calculated by using the one-point density Eq.~(\ref{eq:one}) as
\begin{eqnarray}
\mathbb{E}_{h}\!\left[T_{\text{P}}\right]&=&m\int_{0}^{\infty}\!\!x^{2}h_{1}(x)\dd x \\
&=&-\frac{1}{2}\int_{0}^{\infty}\!\!x^{2}\int_{1}^{\infty}G_{2,3}^{1,1}(\alpha|tx)G_{2,3}^{2,1}(\alpha|tx)\dd t \dd x \nonumber \\
&&-\frac{1}{2}\int_{0}^{\infty}\!\!x^{2}\int_{1}^{\infty}G_{2,3}^{1,1}(\alpha+1|tx)G_{2,3}^{2,1}(\alpha+1|tx)\dd t \dd x, \nonumber
\end{eqnarray}
where we used the representation Eq.~(\ref{eq:01}) instead of Eq.~(\ref{eq:GG}). By changing the order of integration, we arrive at (cf. Eq.~(\ref{eq:Ib}))
\begin{eqnarray}
\mathbb{E}_{h}\!\left[T_{\text{P}}\right]&=&-\frac{1}{2}\int_{1}^{\infty}\left(I_{\alpha}^{(2)}(t)+I_{\alpha+1}^{(2)}(t)\right)\dd t \label{eq:CoI}\\
&=&\frac{m(m+2\alpha+1)}{4(2m+2\alpha+1)}\big(5m^2+10\alpha m+5m+4\alpha^2+4\alpha+2\big), \label{eq:TPm}
\end{eqnarray}
where the last step was obtained by using Eq.~(\ref{eq:b2}) and the fact that
\begin{equation}
\int_{1}^{\infty}\frac{1}{t^{3}}\dd t=\frac{1}{2}.
\end{equation}
The change of the order of integration is justified since the integrals in Eq.~(\ref{eq:CoI}) exist as a result of using the representation Eq.~(\ref{eq:01}). Inserting Eq.~(\ref{eq:TPm}) into Eq.~(\ref{eq:mc3}), one obtains
\begin{equation}\label{eq:aPm}
\mathbb{E}_{f}\!\left[S_{\text{P}}\right]=\frac{5m^2+10\alpha m+5m+4\alpha^2+4\alpha+2}{(2m+2\alpha+1)\left(m^2+2\alpha m+m+2\right)}.
\end{equation}
Finally, evaluating the above expression with the value of $\alpha$ in Eq.~(\ref{eq:aBH}) of the Bures-Hall ensemble, we prove the conjectured formula Eq.~(\ref{eq:Pm}).

\subsection{Average von Neumann Entropy}
Similarly to the steps that have led to Eq.~(\ref{eq:CoI}), the mean value $\mathbb{E}_{h}\!\left[T_{\text{vN}}\right]$ is calculated via the relations Eqs.~(\ref{eq:IH}) and~(\ref{eq:IH0}) as
\begin{eqnarray}
\mathbb{E}_{h}\!\left[T_{\text{vN}}\right]&=&-\frac{1}{2}\int_{1}^{\infty}\left(H_{\alpha}(t)+H_{\alpha+1}(t)\right)\dd t\\
&=&-\frac{1}{2}\left(H_{\alpha}+H_{\alpha+1}\right)\int_{1}^{\infty}\frac{1}{t^{2}}\dd t +\frac{1}{2}\int_{1}^{\infty}\left(I_{\alpha}^{(1)}(t)+I_{\alpha+1}^{(1)}(t)\right)\ln t\dd t.
\end{eqnarray}
The results Eqs.~(\ref{eq:IHf}) and~(\ref{eq:b1}) give us
\begin{eqnarray*}
&&H_{\alpha}+H_{\alpha+1}=-m(m+2\alpha+1)(\psi_{0}(m+\alpha+1)+1),\\
&&I_{\alpha}^{(1)}(t)+I_{\alpha+1}^{(1)}(t)=-m(m+2\alpha+1)t^{-2},
\end{eqnarray*}
and together with the fact that
\begin{equation}
\int_{1}^{\infty}\frac{1}{t^{2}}\dd t=1,~~~~~~\int_{1}^{\infty}\frac{\ln t}{t^{2}}\dd t=1,
\end{equation}
one arrives at
\begin{equation}\label{eq:TavNm}
\mathbb{E}_{h}\!\left[T_{\text{vN}}\right]=\frac{m(m+2\alpha+1)}{2}\psi_{0}(m+\alpha+1).
\end{equation}
Inserting the above result into the moment relation Eq.~(\ref{eq:mc2}), we finally obtain
\begin{equation}\label{eq:avNm}
\mathbb{E}_{f}\!\left[S_{\text{vN}}\right]=\psi_{0}\!\left(\frac{m(m+1)}{2}+\alpha m+1\!\right)-\psi_{0}(m+\alpha+1),
\end{equation}
which upon evaluated at the value of $\alpha$ in Eq.~(\ref{eq:aBH}) proves the conjectured formula Eq.~(\ref{eq:vNm}).


\providecommand{\noopsort}[1]{}\providecommand{\singleletter}[1]{#1}%

\end{document}